\documentclass[twocolumn,amsmath,amssymb,aps,pra,10pt,showpacs]{revtex4}

\usepackage{graphicx}
\usepackage{amsmath}
\usepackage[dvips]{color}

\begin{document}

\title{Principles of thermal design with nematic liquid crystals}

\author{S. Fumeron}
\email{sebastien.fumeron@univ-lorraine.fr}
\affiliation{Laboratoire d'\'Energ\'etique et de M\'ecanique Th\'eorique et Appliqu\'ee, 
CNRS UMR 7563, Nancy Universit\'e, 54506, Vand\oe{uvre} Cedex, France.}
\author{E. Pereira}
\affiliation{Instituto de F\'{i}sica, Universidade Federal de Alagoas, Campus A.C. Sim\~{o}es, 57072-900,
Macei\'{o}, AL, Brazil.}
\author{F. Moraes}
\affiliation{Departamento de F\'{\i}sica, CCEN,  Universidade Federal da Para\'{\i}ba, 58051-970, Caixa Postal
5008, Jo\~ao Pessoa, PB, Brazil.}
\begin{abstract}
Highly engineered materials are arousing great interest because of their ability to manipulate heat, as described by coordinate transformation approach. Based on the recently developed analog gravity models, this paper presents how a simple device based on nematic liquid crystal can achieve in principle either thermal concentration or expulsion. These outcomings are shown to stem from topological properties of a disclination-like structure, induced in the nematic by anchoring conditions.
\end{abstract}


\maketitle

Manipulation of heat flux is currently stimulating intensive research efforts because of its abundant wealth of potential applications. Most challenging stakes are related to thermal shielding/stealth of objects, concentrated photovoltaics (higher conversion efficiency, reduction in the cell area \cite{swanson}) or thermal information processing (heat-flux modulators \cite{ben}, thermal diodes \cite{terrano, li1, li2, zettl}, thermal transistors \cite{li3, saira}, thermal memories\cite{li4,li5}). These prospects come from the possibility to design energy paths in a similar fashion to light in transformation optics \cite{pendry}: the use of nanostructured devices is indeed expected to allow concentrating, shielding and inverting of conductive heat flux \cite{narayana, Guenneau2012, maldovan}. Recent works brought experimental evidences of their efficiency \cite{narayana}. However, besides the technical issues in building such composite materials, they also suffer from a lack of flexibility: for example, once the device has been made, it can shield a region from heat flux, but that function cannot be switch off (which is of prime interest for thermal logic gates or heat flux modulators). 
    
\begin{figure}[htb]
		\includegraphics[height=0.65in]{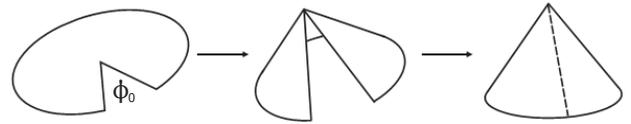}\caption{Volterra process: an angular
    sector $\phi_0$ is removed from a flat surface, the remaining edges are glued together, which results in a conical surface.}\label{cut}
\end{figure}

A promising way to overcome this difficulty relies in the use of nematic liquid crystals (NLC) devices instead of multilayered devices. NLC are generally organic compounds formed by an assembly of rod-like molecules. These latter often consist in a rigid core made of two benzene rings (responsible for crystal-like properties) associated with flexible exterior chains (responsible for fluidity). In other words, molecules in the nematic phase show no positional order (as in ordinary liquids), but orientational order (anisotropy of dielectric permittivities, elastic constants \cite{gennes}..). Thus, NLC are invariant under the symmetry group $D_{\infty}=SO(2)\times Z_2$, that consists in rotations about the molecular axis (called the director) and rotations through $180^{\circ}$ about axes in the orthogonal plane. 
Generally, the alignment of the director exhibits discontinuities: among them, disclinations (also called wedge dislocations) are elastic defects belonging to the first homotopy group $\pi_1$ and they appear as a consequence of $SO(2)$ symmetry-breaking. An intuitive way to understand it is to perform a "cut and glue" Volterra process \cite{kleman2}: the disclination is generated by removing a wedge of material of dihedral angle $\phi_0$  and  gluing the loose faces together (see Fig. \ref{cut} for the two-dimensional case). This generates a positive  curvature disclination. The inverse procedure, inserting a wedge of material, produces a negative curvature disclination (not shown).

Topological defects are the core of fruitful analogies between condensed matter physics and cosmology\cite{chuang, moraes1}. They are elegantly described in terms of Riemann-Cartan geometry \cite{Kondo, Bilby, katanaev1} and therefore they can be studied by using techniques borrowed from General Relativity ($U_{4}$ theory of gravitation \cite{Hehl}): this is the core of so-called analog gravity. Compared to ordinary elasticity theory (OET), this approach has two major assets : first, its accuracy (OET only reproduces the first-order approximation of the geometric theory of defects \cite{katanaev3}) and second its versatility (changing the kind of defect only requires to change the metric, instead of a complicated set of boundary conditions in OET). In this framework, a disclination in NLC acts analogously to a gravitational line source and the medium surrounding is described by the metric of a global cosmic string \cite{ermseb1, Vilenkin}: 
\begin{eqnarray}
	ds^2=dr^2+\alpha^2r^2d\phi^2+dz^2,\label{desclids}
\end{eqnarray}
which describes a conical geometry. For both cosmic strings and disclinations in elastic solids, the parameter $\alpha$ is related to the sector removed/added by $\phi_0=2\pi(1-\alpha)$, where $\alpha<1$ represents a removal of matter, while $\alpha>1$ represents an addition of matter.

In cartesian coordinates, the covariant components of the metric (\ref{desclids}) are thus given by:
\begin{eqnarray}
g_{ij}=\left( 
\begin{array}{ccc}
\frac{x^2+\alpha^2 y^2}{r^2}&\frac{xy(1-\alpha^2)}{r^2}&0\\
\frac{xy(1-\alpha^2)}{r^2}&\frac{\alpha^2 x^2+y^2}{r^2}&0\\
0&0&1\\
\end{array}
\right)
\label{desclimetric-cart}
\end{eqnarray}

\begin{figure}[htb]
		\includegraphics[height=1.8in]{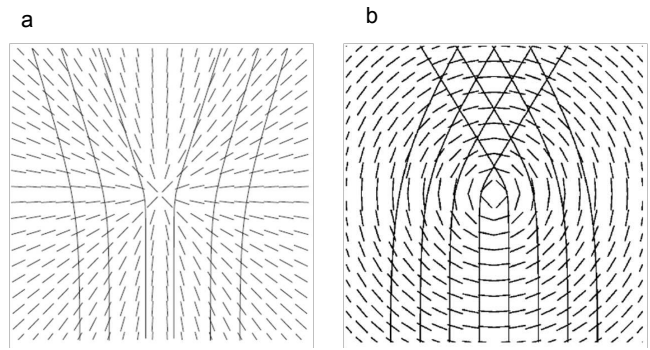}\caption{Director field (dotted lines) and geodesics (plain lines) in the presence of a disclination line. (a) Case $\alpha > 1$. (b) Case $\alpha' < 1$. If it is the same material, $\alpha' = 1/ \alpha$.}\label{conic}
\end{figure}

In the diffuse regime, heat conduction in the nematic phase is modified by the presence of topological defects. For a given metric $g_{ij}$, heat equations governing conductive flux and temperature fields are given by \cite{fumeron1}:
\begin{eqnarray}
&q^i=-\lambda^{ij} \partial_j T  ,\\
&div \textbf{\textit{q}}=p ,
\end{eqnarray} 
with
\begin{eqnarray}
&&\lambda^{ij}=\lambda\; g^{ij} ,\label{conduc-general} \\
&&p=-\frac{\lambda}{2} g^{ij}\; \partial_j T \;\partial_i \ln{g} ,\label{heat-general} 
\end{eqnarray} 
where $g=\det g_{ij}$ and $\lambda$ is the thermal conductivity of the isotropic liquid phase. Generally speaking, it appears that a non-trivial metric modifies the thermal conductivity. Moreover, it can also introduce an effective internal heat source ($p>0$) or sink ($p<0$) that is directly coupled to the temperature fields. In the case of (\ref{desclimetric-cart}), heat conduction locally occurs as in a monoclinic crystal \cite{fumeron1} and as $g=\alpha^2=const$, the disclination does not introduce any internal source term. Similarly to matter in cosmic strings wakes \cite{kibble}, heat flux lines stay in planes $z=const.$ but feel the curvature of the background conical geometry generated by the disclination (see Fig. \ref{conic}). 
As the Riemann tensor is null everywhere, except at the defect core where it has a $\delta$-function singularity given by \cite{sokolov77}
\begin{equation}
	R_{12}^{12}=2\pi\frac{1-\alpha}{\alpha}\delta(r),
\end{equation} 
the distortion of flux lines when approaching the defect core also testifies that disclinations are responsible for an elastic analog of Aharonov-Bohm effect (as already noticed in \cite{azevedo2, carvalhoah}).

\begin{figure}[h!]
\includegraphics[height=2.5in]{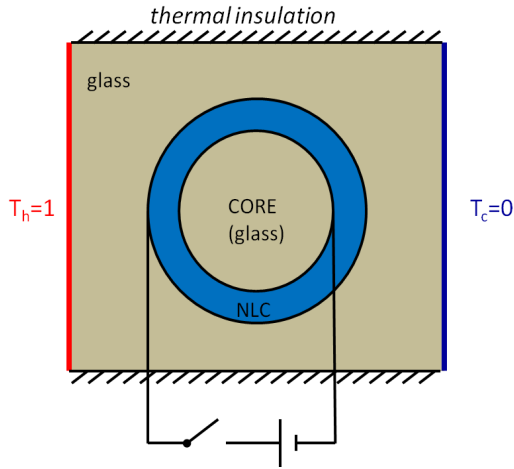}\caption{Principle of the thermal control device}\label{device}
\end{figure}

Now that the specific features of heat conduction near a disclination have been reviewed, we are investigating the possibility to implement them for thermal design. A qualitative understanding of it can be obtained from a parametric study of the following configuration: a hollow cylinder, inside which there is the core region where one aims at controlling the conductive heat flux, is inserted inside a conducting solid sandwiched between two heated vertical plates (see figure \ref{device}). In normalized units, cold temperature is set to $T_c=0$, whereas hot temperature is set to $T_h=1$. The host material consists in an homogenous isotropic medium with unit thermal conductivity, whereas the thick cylinder consists in a NLC for which the spatial configuration of the director $\hat{n}$ is that of a disclination. We can think of a nematic director field configuration like the one in Fig. 2a without the singularity on its axis, by isolating the axis with a cylindrical wall. In order to have the same radial orientation of the director field,  both inner and outer boundaries can be prepared to provide homeotropic anchoring to the nematic molecules. As there is no disclination core, such configuration is topologically stable. That is, no escape in the third dimension can occur. The stability of the mesophase is more problematic. Indeed, liquid crystals made from a single kind of organic molecules are thermotropic and in practice, they exhibit a nematic phase in a very narrow range of temperatures (typically a few tens of degrees). But for thermal management, NLC with low melting and high clearing temperatures are required. This can be achieved by using eutectic liquid crystals mixtures (or ``guest-host systems"): by adjusting the proportions of each compound (Schroeder-Van Laar law), the nematic range can be larger than $100$ K\cite{eutectic,pieranski}.

The thermal conductivity of the isotropic fluid phase is taken equal to that of the host medium, to avoid additional thermal effects (such as Kapitza resistance). We will consider three cases: 1) the reference case consisting in a NLC without disclination ($\alpha=1)$; 2) a strength +1 disclination with the director aligned radially, for which $\alpha=\sqrt{C_{33}/C_{11}}=2$ (Fig. 2a);  3) again a strength +1 disclination but now with the director circling the defect axis, for which $\alpha=\sqrt{C_{11}/C_{33}}=0.5$ (Fig. 2b). In each case, the defect axis coincides with the axis of the cylinder. Based on COMSOL Multiphysics finite element-based simulations, we first present results related to $\alpha=1$.
\begin{figure}[h!]
\includegraphics[height=1.55in]{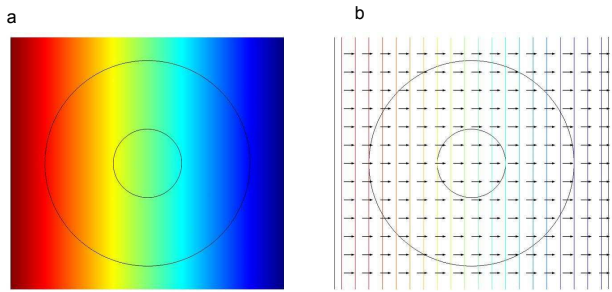}\caption{Reference case without the device ($\alpha=1$). (a) Temperature field. (b) Heat flux lines.}\label{referent}
\end{figure}
Temperature fields and heat flux lines in Fig. \ref{referent} are the regular solutions of steady-state heat conduction equations inside isotropic solids: heat flux lines follow the direction of the temperature gradient isothermal lines ($x$-axis), whereas isothermal lines correspond to the planes $y=const$ in agreement with boundary conditions. 

\begin{figure}[h!]
\includegraphics[height=1.55in]{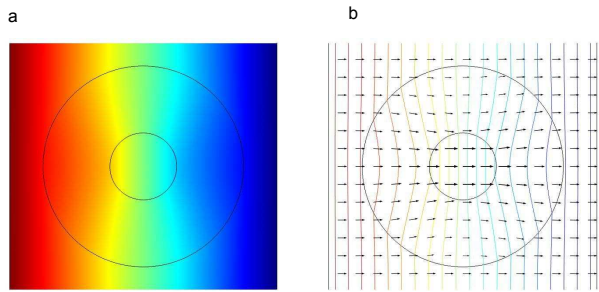}\caption{Concentrator configuration with $\alpha=2$. (a) Temperature field. (b) Heat flux lines.}\label{concentrator}
\end{figure}

Now, let us examine on Figure \ref{concentrator} the case worresponding to $\alpha=2$.
Compared to previous case, isothermal lines bend in order to enter the core of the device. Besides, heat flux lines are also converging inside the inner region, such that the norm of heat flux vectors is globally increasing in this area. Therefore, the device is responsible for focusing conductive heat inside the core, which can be thought of as a heat concentration phenomenon.   

\begin{figure}[h!]
		\includegraphics[height=1.55in]{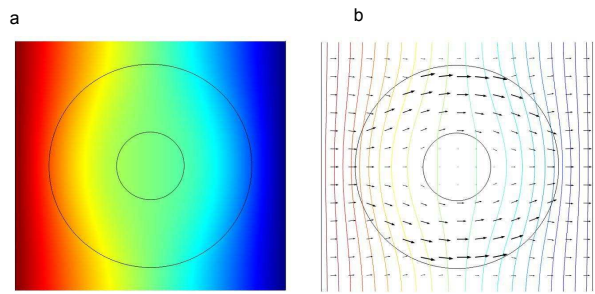}\caption{Repeller configuration with $\alpha=0.5$. (a) Temperature field. (b) Heat flux lines.}\label{cloak}
\end{figure}

Now, we consider on Figure \ref{cloak} the case $\alpha=0.5$ for which the device behaves as a heat repeller. Indeed, isothermal lines are diverted from the inner region. Moreover, in a similar fashion to light paths in optical cloaking\cite{pendry}, heat flux lines are expelled from the core and the norm of heat flux vectors is vanishing in this area. However, we do not expect the possibility of perfect heat repulsion (null heat flux in the core), as that would imply a thermal singularity related to the heat flux line headed directly towards the centre of the device (heat does not know in which direction to deviate from the inner region).

It must be emphasized that these two behaviours (repelling and concentrating) cannot be understood in terms of geodesics of the effective metric (\ref{desclimetric-cart}), as it is usually the case in analog gravity models. In general relativity, geodesics are the paths followed by free-falling particles in a curved spacetime. They are the solutions of Euler-Lagrange equations, obtained by varying the Einstein-Hilbert action, and they are a generalization of Newton's second law $m\:\bold{a}=\bold{F}$ in the case where $\bold{F}=0$ (uniform motion). Therefore, considering the presence of a temperature gradient is equivalent to add a thermal constraint to Euler-Lagrange equations. Another way of understanding it is to take a mechanical standpoint. Heat conduction can be thought of as a phonon gas, and as such, it obeys fluid mechanics. In this frame, the temperature gradient acts as a driving force for the heat flow (see \cite{Cao} for temperature gradient as an effective pressure in phonon hydrodynamics), so that the heat flux cannot be considered as free-falling.

\begin{figure}
		\includegraphics[height=1.9in]{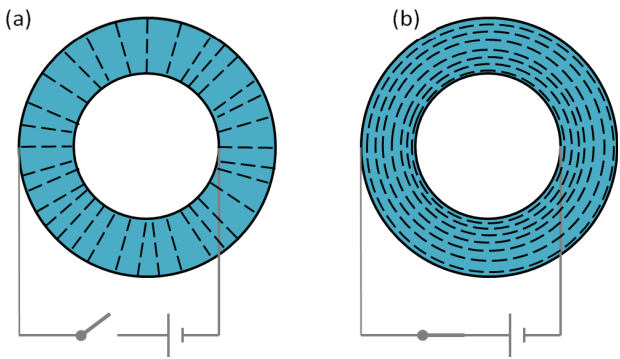}\caption{Principle of concentrator-repeller switching based on bistable anchoring transition. (a) Homeotropic anchoring in the absence of external field. (b) Transition to planar anchoring due to electric field. This latter is shut off after transition is performed.}\label{switch}
\end{figure}

As previously mentioned, the main asset of our device over systems based on composite materials relies in its versatility. As a matter of fact, the concentrator is achieved for a NLC such that $C_{33}>C_{11}$ ($\alpha>1$), with director normal to the boundaries ($\hat{n}=\hat{r}$). Conversely, the cloak configuration may be achieved for  the same NLC, but with the director parallel to the boundaries ($\hat{n}=\hat{\phi}$) such that $\alpha>1$. Therefore, rotating the rod-like molecules by $90$ degrees enables to switch at will from the concentrator to the cloak. This can be achieved by electric-field-driven bistable anchoring (see figure \ref{switch}). Anchoring yields a constant orientation of the thermotropic NLC molecules due to particular surface conditions. In the presence of a surfactant-treated surface (for example silane), rods align normally to the surface and anchoring is said homeotropic. On the contrary, in the presence of a polymer coating such as PVA (polyvinyl alcohol), anchoring is planar and calamitic molecules align parallel to the surface. One technical issue to develop this device is to anchor NLC molecules on curved substrates. Usual photopolymer-based techniques require a well-defined angle of incidence for light, which make them ill-adapted for non-planar surfaces. Gupta et al. \cite{Gupta} performed NLC anchoring on curved surfaces by using self-assembled monolayers (SAM) formed from alkanethiols. Two additional assets for SAM-based anchoring are that 1) they are stable upon application of the electric field, and 2) polymerizable SAM are expected to be chemically stable over years (no oxidative degradation). Above the Frederiks transition, application of an electric field leads to unstable new states of orientation (depending on the sign of its dielectric anisotropy)\cite{gennes}: as soon as the field is removed, molecules relax back to the original orientation fixed by anchoring conditions. However, for dye-doped NLC, sufficiently high values of the field were shown to induce stable anchoring transitions between homeotropic and planar states: back switching does not occur even when the field is off\cite{kim-bistab}. In principle, such effect can be used to switch between the concentrator and the repeller by applying during a short time an electrical potential difference between both sides of the hollow cylinder. As the electrodes are turned on only during the time required for the anchoring transition to occur, electric-field induced instabilities do not appear when the device is working. 

Both configurations (concentrator and repeller) have been considered with horizontal temperature gradients, so that no Rayleigh-B$\acute{\text{e}}$rnard convection can occur. However, as the thermal gradient is not parallel to the gravity field $\textbf{g}$, weak thermoconvective instabilities can develop in the annulus domain. These instabilities vanish provided that the device is thin enough. Although an accurate justification requires solving the equations of nematohydrodynamics, this effect can be understood qualitatively from the usual viscous dissipation power \cite{LandauMF}. For an annulus region of thickness $d$, with internal radius $r_i$ and outer radius $r_e$, consider a thermoconvective instability with mean velocity $U$. Therefore, the dissipated power can roughly be expressed as a scaling law according to :
\begin{eqnarray}
P_{dis}&=&2\eta\iiint \Bar{\Bar{D}}:\Bar{\Bar{D}}\: d^3x\sim \eta\frac{U^2}{d}\pi(r_e^2-r_i^2),
\end{eqnarray} 
where $\eta$ is the mean shear viscosity of the nematic and $\Bar{\Bar{D}}$ is the strain rate tensor. Therefore, inner viscous dissipation is maximized in a disc-like device (low thickness $d$ and small core regions $r_i$) and it is also expected to be enhanced by dissipation at boundaries (anchoring conditions).

In this work, we examined the possibility of thermal management from a liquid crystal-based device. It relies on reproducing inside a hollow cylinder the topology of a disclination, which can imprint curvature to conductive flux lines. Numerical results confirm the possibility of heat guiding phenomena: in particular, we showed that the device can be tuned to either concentrate or expulse heat flux. The switching frequency is limited by relaxation times governing the anchoring transition (several milliseconds for dye-doped nematics \cite{kim-bistab}). In principle, it can be designed to work in a relatively large range of temperatures. Concentrator or repeller configurations are both mechanically stable and thermal instabilites can be neglected by adequate choice of geometry (disc-like). However, it is mandatory for practical developments of such device to quantitatively study these aspects through optimal choice of materials. This will be considered in our next works.

E. Pereira thanks financial support from Brazilian agencies FAPEAL and CNPq. F. Moraes thanks financial support from INCT-FCx, CNPq and CAPES.



\begin{thebibliography}{0}
\expandafter\ifx\csname natexlab\endcsname\relax\def\natexlab#1{#1}\fi
\expandafter\ifx\csname bibnamefont\endcsname\relax
  \def\bibnamefont#1{#1}\fi
\expandafter\ifx\csname bibfnamefont\endcsname\relax
  \def\bibfnamefont#1{#1}\fi
\expandafter\ifx\csname citenamefont\endcsname\relax
  \def\citenamefont#1{#1}\fi
\expandafter\ifx\csname url\endcsname\relax
  \def\url#1{\texttt{#1}}\fi
\expandafter\ifx\csname urlprefix\endcsname\relax\def\urlprefix{URL }\fi
\providecommand{\bibinfo}[2]{#2}
\providecommand{\eprint}[2][]{\url{#2}}

\end{thebibliography}


\begin{thebibliography}{100}

\bibitem{ben}P.J. van Zwol and K. Joulain and P. Ben Abdallah and J.J. Greffet and J. Chevrier, Phys. Rev. B \textbf{83}, 201404 (2011).
\bibitem{terrano}M. Terrano and M. Peyrard and G. Casati, Phys. Rev. Lett. \textbf{88}, 094302 (2002). 
\bibitem{li1}B. Li and L. Wang and G. Casati, Phys. Rev. Lett. \textbf{93}, 184301(2004). 
\bibitem{li2}B. Li and J. Lan and L. Wang, Phys. Rev. Lett. \textbf{95}, 104302 (2005). 
\bibitem{zettl}C. W. Chang and D. Okawa and A. Majumdar and A. Zettl, Science \textbf{314}, 1121 (2006). 
\bibitem{li3}B. Li and L. Wang and G. Casati, Appl. Phys. Lett. \textbf{88}, 143501 (2006). 
\bibitem{saira}O. P. Saira and M. Meschke and F. Giazotto and A. M. Savin and M. Mottonen and J. P. Pekola, Phys. Rev. Lett. \textbf{99}, 027203 (2007). 
\bibitem{li4}L. Wang and B. Li, Phys. Rev. Lett. \textbf{99}, 177208 (2007). 
\bibitem{li5}L. Wang and B. Li, Phys. Rev. Lett. \textbf{101}, 267203 (2008). 
\bibitem{pendry}J.B. Pendry and D. Schurig and D.R. Smith, Science \textbf{312}, 1780 (2006). 
\bibitem{narayana}S. Narayana and Y. Sato, Phys. Rev. Lett.  \textbf{108}, 214303 (2012). 
\bibitem{Guenneau2012}S. Guenneau and C. Amra and D. Veynante, Opt. Exp. \textbf{20}, 8207 (2012). 
\bibitem{maldovan}M. Maldovan , Phys. Rev. Lett. \textbf{110}, 025902 (2013). 
\bibitem{gennes}P. G. de Gennes and J. Prost, \emph{The Physics of Liquid Crystals} (Claredon Press, Oxford, 1992), 2nd ed.
\bibitem{kleman2}M. Kleman and J. Friedel , Rev. Mod. Phys. \textbf{80}, 61 (2008). 
\bibitem{chuang} I. Chuang and B. Yurke and R. Durrer and N. Turok, Science \textbf{251}, 1336 (1991). 
\bibitem{moraes1} C. Furtado and F. Moraes and A. de M. Carvalho, Phys. Lett. A \textbf{372}, 5368 (2008). 
\bibitem{Kondo} K. Kondo, Jpn. Nat. Congr. Appl. Mech.: Proc. (Tokyo) \textbf{2}, 41 (1952). 
\bibitem{Bilby} B. A. Bilby and R. Bullough and E. Smith, Proc. R. Soc. A \textbf{231}, 263 (1955). 
\bibitem{katanaev1} M. O. Katanaev and I. V. Volovich, Ann. Phys. \textbf{216}, 1 (1992). 
\bibitem{hehl} F.W. Hehl and P. Von Der Heyde and G.D. Kerlick and J.N. Nester, Rev. Mod. Phys. \textbf{48}, 393 (1976). 
\bibitem{katanaev3} M. O. Katanaev, Phys.-Usp \textbf{48}, 675 (2005). 
\bibitem{ermseb1} E. Pereira and S. Fumeron and F. Moraes, Phys. Rev. E \textbf{87}, 022506, 049904 (2013). 
\bibitem{vilenkin} A. Vilenkin, Phys. Rev. D \textbf{23}, 852 (1981). 
\bibitem{caio2} C S\'atiro and F. Moraes, Eur. Phys. J. E \textbf{25}, 425 (2008). 
\bibitem{fumeron1} S. Fumeron and E. Pereira and F. Moraes , Int. J. Therm. Sci. \textbf{67}, 64 (2013). 
\bibitem{kibble} M.B. Hindmarsh and T.W.B. Kibble , Rep. Prog. Phys. \textbf{58}, 477 (1995). 
\bibitem{sokolov77} D. Sokolov and A. Starobinskii , Sov. Phys. Dokl. \textbf{22}, 312 (1977). 
\bibitem{azevedo2} S. Azevedo and F. Moraes , Phys. Lett. A \textbf{246}, 374 (1998). 
\bibitem{carvalhoah} C. Furtado and A. M. de Carvalho and C. A. de Lima Ribeiro ,  Mod. Phys. Lett. A \textbf{21}, 1393 (2006). 
\bibitem{eutectic} S.T. Wu and C.S. Hsu and K.F. Shyul ,  Appl. Phys. Lett. \textbf{74}, 344 (1999). 
\bibitem{pieranski}P. Oswald and P. Pieranski, \emph{Nematic and Cholesteric Liquid Crystals: Concepts and Physical Properties Illustrated by Experiments} (CRC Press, 2005), 1st ed.
\bibitem{kim-bistab} J.K. Kim and K.V. Le and S. Dhara and F. Araoka and K. Ishikawa and H. Takezoe ,  J. Appl. Phys. \textbf{107}, 123108 (2010). 
\bibitem{landauMF} L.D. Landau and E. Lifschitz, \emph{Fluid mechanics}  (Butterworth-Heinemann, New York, 1987), 2nd
ed.


\end{thebibliography}

\end{document}